# $\Gamma_5$ quasiparticles and avoided quantum criticality in U(Ru,Rh)$_2$Si$_2$

A. V. Silhanek[a], N. Harrison[a], C. D. Batista[b], M. Jaime[a], A. Lacerda[a], H. Amitsuka[c], and J. A. Mydosh[d]

[a]*Los Alamos National Laboratory, MS E536, Los Alamos, NM 87545, USA*

[b]*Los Alamos National Laboratory, MS B262, Los Alamos, NM 87545, USA*

[c]*Graduate School of Science, Hokkaido University, N10W8, Sapporo 060-0810, Japan*

[d]*Max-Planck-Insitut für Chemische Physik fester Stoffe, 01187 Dresden, Germany*

**Abstract**

We discuss recent specific heat data in high magnetic fields on URu$_2$Si$_2$ and 4% Rh-doped URu$_2$Si$_2$ as well as previously published de Haas-van Alphen data at lower magnetic fields in pure URu$_2$Si$_2$; both of which are consistent with quasiparticle bands formed from a hybridization between 5$f$-electron $\Gamma_5$ doublets and regular conduction electrons. The system exhibits itinerant electron metamagnetism that gives rise to a putative quantum critical point at ~ 34 -37 T (depending on the % of Rh) that is subsequently unstable to field-induced phases.

*Keywords:* U(Ru,Rh)$_2$Si$_2$; quasiparticles; quantum criticality; metamagnetism; electric quadrupoles

URu$_2$Si$_2$ is a U-based heavy fermion compound that exhibits an unidentified 'hidden order' (HO) phase at temperatures below $T_o$ = 17.5 K [1,2]. In 2002, Park *et al.* [3] published the results of a high energy inelastic neutron scattering study in which they identified an intermultiplet transition at ~ 360 meV, thus providing rather compelling evidence for U$^{4+}$ and a 5$f^2$ electronic configuration. On combining this experimental finding with prior de Haas-van Alphen (dHvA) data published by Ohkuni *et al.* in 1999 [4], we can conclude that URu$_2$Si$_2$ possesses a heavy Fermi liquid composed of $\Gamma_5$ quasiparticles. Such quasiparticles come into existence following a hybridization between the conduction electrons and 5$f^2$ magnetic doublets, as can be described using the Anderson lattice model [5]. One consequence of such a hybridization is that the itinerant quasiparticles are strongly affected by spin-orbit coupling, adopting the spin and orbital quantum numbers of the lowest energy local crystal electric field level of the 2 5$f$-electrons. The dHvA data therefore prove that the lowest crystal field level is a pure Ising 5$f^2$ doublet, of which there only exist non-Kramers doublets having the $\Gamma_5$ orbital manifold.

The published data of interest displays nodes in the magnetic field-orientation dependent dHvA amplitude (see Fig. 18 of Ref. [4]). Such nodes are the consequence of the product $m^*g$ of the orbitally averaged effective mass (in units of the free electron mass $m_e$) and the effective $g$-factor (for a pseudospin of ½) changing with the orientation of the magnetic field $H$, with a node occurring each time $m^*g$ is equal to an odd integer. Filled circles in Fig. 1a correspond to the odd values of $m^*g$ extracted from the nodes in the published data [4]. The existence of as many as 16 nodes is exceptional, given the relatively small change in the effective mass with angle, presented in Fig. 1b. Open circles in Fig. 1a show the values of $m^*g$ expected for $g = g_z\cos\theta$, as for a pure Ising spin with $g_z$ = 2.6 (close to the theoretical estimate of $g_z$ = 2.4), and $m^*$ interpolated using a spline fit from Fig. 1b.

The existence of quasiparticle bands composed of two highly anisotropic spin degrees of freedom implies that one can make comparisons with other highly anisotropic Fermi liquid materials, such as CeRu$_2$Si$_2$. Indeed, both CeRu$_2$Si$_2$ and URu$_2$Si$_2$ display itinerant electron metamagnetism [5,6,7], with their $T$-dependent specific heats becoming nearly identical at magnetic fields above the metamagnetic transition field. Only at low magnetic fields are they significantly different, with $URu_2Si_2$ having a lower carrier density. This is partly due to the bandstructure yielding smaller closed sections of Fermi surface [4] and partly due to the gapping of much of the Fermi surface within the HO phase; both giving rise to a lower electronic contribution to the specific heat as $T \rightarrow 0$. Superexchange interactions are also far more significant for U than for Ce, thereby increasing the likelihood of ordered phases.

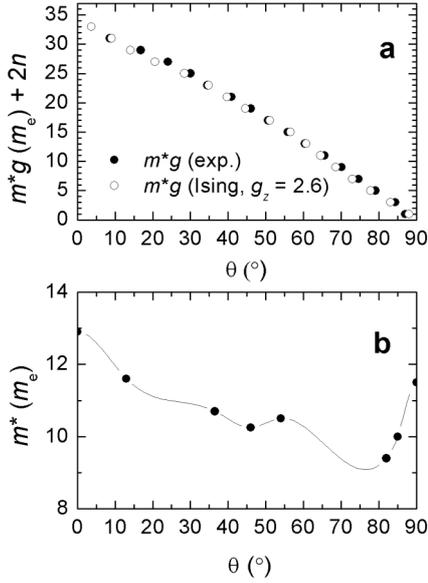

Fig. 1. (a) Comparison of the experimentally determined product $m^*g$ (filled circles) with that predicted (open circles) according to the measured mass as a function of angle shown in (b) and an Ising $g$-factor as described in the text, both plotted as a function of the angle θ between the magnetic field and the c-axis.

Figure 2 shows an intensity plot of the quasiparticle density of states (DOS) calculated using a model that is able to explain the $T$-dependent specific heat data at magnetic fields above $\mu_0 H_m \sim 34.4$ T in 4% Rh-doped $URu_2Si_2$ [7]; the specific heat is very similar to that of pure $URu_2Si_2$ at these fields. The quasiparticle band model is fitted to the data for $\mu_0 H > 38$ T, where the magnetic field causes the separation of the pseudospin up and pseuodospin down bands to be strongly field dependent. The same is also true for the bandwidth, which becomes progressively more narrow as $H \rightarrow H_m$, therefore being consistent with a putative quantum critical point at $H_m$. The intensity plot for $\mu_0 H < 38$ T is based on a linear extrapolation of specific heat fitting parameters from 38 T and above [7]. Rather than a quantum critical point actually being realized at $H_m$, a first order phase transition takes place at ~38 T into a field-induced ordered phase [7]. The resultant form of the density of electronic states within the ordered phase is unknown, but we should expect a gap (or pseudogap) of order 3 meV or higher to open up around the Fermi energy (if we naively apply BCS theory to the optimal ordering temperature [7]).

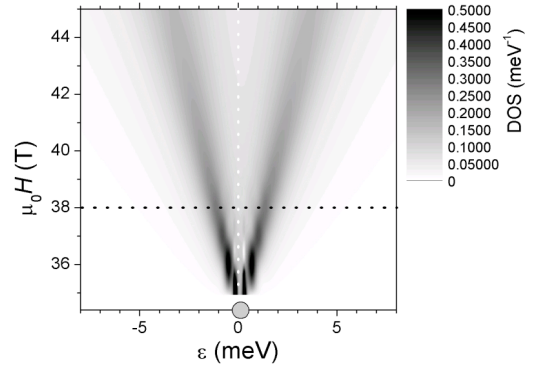

Fig. 2. Intensity plot of the DOS of the quasiparticle bands able to reproduce specific heat data [7]. The black and while dotted lines correspond to the first order transition field and the Fermi energy respectively.

The quasiparticles in $URu_2Si_2$ (and Rh-doped $URu_2Si_2$) are unique in that they acquire electric quadrupolar as well as Ising spin of freedom as a consequence of hybridization. Superexchange interactions can therefore lead to competing electric quadrupolar and Ising antiferromagnetic ordered groundstates, with the former becoming increasingly favorable in strong magnetic fields. Since the former involves charge degrees of freedom, it may be more strongly coupled to the crystal lattice, possibly explaining why the phase transitions become first order at low $T$.

Returning now to general question of order parameters in pure $URu_2Si_2$ for $H < H_m$, it is now rather clear that theoretical models for the HO phase must take into consideration the fate of the electric quadrupolar and Ising degrees of freedom of the itinerant $\Gamma_5$ quasiparticles. Antiferroquadrupolar order has been proposed, although only from a purely local moment perspective [8].